\documentstyle[preprint,aps,prl]{revtex}
\def\ltap{\ \raise.3ex\hbox{$<$\kern-.75em\lower1ex\hbox{$\sim$}}\ }
\def\gtap{\ \raise.3ex\hbox{$>$\kern-.75em\lower1ex\hbox{$\sim$}}\ }

\def\be{\begin{equation}}
\def\ee{\end{equation}}
\begin{document}


\preprint{UW/PT-97-03}

\title{Horizontal, Anomalous $U(1)$ Symmetry for the More Minimal
        Supersymmetric Standard Model}

\author{Ann E.  Nelson and David Wright}

\address
    {%
    Department of Physics, Box 351560,
    University of Washington,
    Seattle, Washington 98195
    {\tt anelson@phys.washington.edu, wright@phys.washington.edu}}%
\date{February 1997}

\maketitle

\begin {abstract}
We construct explicit examples with a
horizontal, ``anomalous'' $U(1)$ gauge group, which, in
a supersymmetric extension of the
standard model, reproduce qualitative features of the fermion spectrum and
 CKM matrix, and   suppress FCNC and proton decay
rates without the imposition of global symmetries.
We review the motivation for such ``more'' minimal supersymmetric
standard models and their predictions for the sparticle spectrum.
There is a  mass hierarchy in the scalar sector  which is   the inverse
of
the fermion mass hierarchy.
We show in detail why $\Delta S=2$
FCNC are greatly suppressed when compared with naive
estimates for
nondegenerate
squarks.
\end {abstract}
\newpage


The minimal supersymmetric standard model (MSSM) with stop-induced electroweak
symmetry breaking naturally stabilizes the electroweak scale, but fails to
exhibit the accidental global symmetries of the standard model (SM) which
inhibit flavor-changing neutral currents (FCNC),
lepton flavor violation (LFV), electric dipole moments (EDMs) and
proton decay.  To agree with experimental bounds,
such processes must be suppressed by imposing
approximate global symmetries to produce degeneracy or alignment
in the squark and slepton mass matrices.
Recently, Cohen, Kaplan, and Nelson \cite{CKN} have proposed an alternative
framework in which the gauginos, higgsinos, and third generation squarks
are sufficiently light to naturally stabilize the electroweak scale,
while the first two generations of
squarks and sleptons are sufficiently heavy to suppress FCNC, LFV, and EDMs
below experimental bounds.
Such models are ``more'' minimally supersymmetric than the MSSM
in the sense that they do not require the ad hoc supposition of
degeneracy or alignment.
The anomalous $U(1)_X$ used by Dvali and Pomarol \cite{DP},
and Binetruy and Dudas \cite{BD}
to break supersymmetry (hereafter DPBD mechanism)
could induce the required mass hierarchy, provided third
generation matter carries no $X$-charge.
In this letter, we show that an anomalous
$U(1)_X$ can also explain the order of magnitude of
fermion masses and mixing angles, and
suppress proton decay from dimension-5 operators.
Our criteria for assigning $U(1)_X$ charges
differ significantly from those considered previously in this context
\cite{BD,MR}.

The DPBD mechanism of supersymmetry breaking requires a
gauged $U(1)_X$ with positively and negatively $X$-charged matter
superfields $P$ and $N$.  If $X$-symmetry is anomalous (i.e. $\rm{tr}(X)
\ne 0$) below some scale $M$, the effective theory below this scale
includes a Fayet-Iliopoulos term $\sim g_X^2 \, \rm{tr} X \, M^2$
\cite{witetc}.
For mathematical consistency, one can either imagine that $M$ is the
Planck scale and the anomaly is canceled via the Green-Schwarz
mechanism \cite{GS}, or that $M$ is some other scale at
which anomaly-canceling matter lives.
If the superpotential also contains a term of the form $m P N$,
introduced either explicitly or dynamically by nonperturbative physics,
the various fields obtain vacuum expectation values
\begin{eqnarray}
        \phi_N  &=& N_1                 = \epsilon M \\
        F_P     &=& P_{\theta \theta}   = \epsilon m M \\
        D_X     &=& (V_X)_{\theta \theta \bar\theta\bar\theta}  = m^2 / g^2
\end{eqnarray}
where $\epsilon$ is a computable number, typically somewhat smaller than
unity\footnote{If we consider the DPBD mechanism in the context of tree-level
supergravity (SUGY) with $M$ set equal to the Planck mass, the vevs
obtained above are perturbed only slightly, $\phi_P$ and consequently
$F_N$ get small vevs, and the gravitino mass $m_{3/2} \sim \epsilon m$.}.

The D-term vev gives each scalar matter field $\phi$ a mass squared
$X_{\phi} m^2$ proportional to its $X$-charge; thus $m$ is a mass scale for
$X$-charged scalar matter.
Our phenomenological philosophy, following ref.~\cite{CKN}, is to set $m$
high relative to a standard model physics scale and move the
first and second generation squarks and sleptons to the scale $m$
by giving their corresponding superfields positive $X$-charge.
A straightforward 1-loop calculation will verify that the
electroweak scale can be reproduced with less than 10\% cancelation
between bare higgs masses squared and 1-loop radiative corrections
only if all particles which couple strongly to the higgses weigh in
at less than $\sim 1$~TeV, the so-called ``'t Hooft bound''.
Since top and left-handed bottom squarks couple strongly to higgses,
naturalness of the electroweak scale requires them to remain
lighter than $\ltap$~1~TeV.  We accomplish
this by assigning the corresponding superfields zero $X$-charge.
Since we do not wish to
break any of the standard model gauge symmetries at a scale $\sim m$, all
chiral superfields charged under SM gauge groups must have nonnegative
$X$-charge.

SM Yukawa couplings appear in the MSSM as terms in the superpotential of
the form $LRH$,
where $L$ and $R$ are  $SU(2)$ doublet and singlet matter superfields,
respectively, and $H$ is the appropriate higgs superfield, either the up-type
$h_u$ or the down-type $h_d$. Most such terms are forbidden by $X$-symmetry
because $LRH$ is not, in general, an $X$-singlet.
However, nonrenormalizable operators, induced in the effective superpotential
by physics at the scale $M$, of the form
\be
        \frac{ \left( L R H \right)_{\theta \theta}
        \left( N^n \right)_1 }{ M^n } =
        \epsilon^n \Psi_L \Psi_R \phi_H
\ee
mimic SM Yukawa couplings because of the expectation value of $\phi_N$.
Here $n$ is a sum of $X$-charges chosen so that $LRHN^n$ is an $X$-singlet.
Assuming such operators are in fact induced with coefficients whose
ratios are $O(1)$, we can, given a set of $X$-charge assignments for
matter superfields, make order-of-magnitude estimates of SM mass matrix
elements
\be
        \lambda^u_{ij} \propto \epsilon^{q_i + \bar u_j+ h_u} \tan \beta \qquad
        \lambda^d_{ij} \propto \epsilon^{q_i + \bar d_j+h_d} \qquad
        \lambda^l_{ij} \propto \epsilon^{\ell_i + \bar e_j+h_d}
\ee
where we have used symbols for the fields to indicate their respective
$X$-charges, and
\be \tan\beta\equiv \langle\phi_{h_u}\rangle/\langle\phi_{h_d}\rangle\ .
\ee
Perturbative diagonalization yields fermion masses and
intergenerational CKM matrix elements
\be
\label{fmass}
        m^u_i \propto \epsilon^{q_i + \bar u_i+ h_u} \tan \beta \qquad
        m^d_i \propto \epsilon^{q_i + \bar d_i+h_d} \qquad
        m^e_i \propto \epsilon^{\ell_i + \bar e_i+h_d} \qquad
        V_{ij} \sim \epsilon^{\left| q_i - q_j \right|}\ .
\ee
Such mass matrix models were introduced by Froggatt and Nielsen \cite{FN}
and have since been explored by many authors \cite{IR}.
The existence of such a $U(1)_X$ horizontal symmetry underlying SM mass
matrices makes the prediction for intergenerational CKM matrix elements
\be
        V_{12} V_{23} \sim V_{13}\ .
\ee
Since $V_{12} \sim \epsilon$, $V_{23} \sim \epsilon^2$, and $V_{13} \sim
\epsilon^3$, for $\epsilon \sim 1/5$, this relation is empirically
confirmed.
In order that the top quark Yukawa coupling be unsuppressed, we must have
$q_3 = \bar{u}_3 = h_u = 0$.  For $\epsilon \sim 1/5$, the CKM matrix
elements are then best explained by $q_1 = 3 , q_2 = 2$.

Squark mass matrix entries are also induced by nonrenormalizable operators
generated by physics at the scale $M$.  In particular, the operator
\be
\label{eq:sqmass}
        \frac{ \left( A^* B \right)_1 \left( N^n \right)_1
        \left( P^* P \right)_{\theta \theta \bar{\theta}
        \bar{\theta}} }{ M^{2+n} } =
        \epsilon^{2+n} m^2 \phi^*_A \phi_B
\ee
with $n = X_B - X_A$ ( or $n = X_A - X_B$ with $N$ replaced by $N^*$ ),
mimics a scalar mass matrix entry.  Note that such
operators give masses of order $\epsilon m$ even
to scalars with vanishing $X$-charge.
Since naturalness requires $\epsilon m \ltap$~1~TeV,
we obtain the constraint $m \ltap$~5~TeV.
Note also that $SU(2)$-invariance requires that $A$ and $B$
in eq.~({\ref{eq:sqmass}) be either both left-handed doublet superfields or
both right-handed singlet superfields; such operators induce no
mixing between left- and right- handed squarks.
Although left-right squark mixing exists in the models we consider, it is
heavily suppressed.
Neglecting left-right mixing, we write four uncoupled squark mass matrices.
\begin{eqnarray}
        \lambda^{uL,dL}_{ij} &\propto& \epsilon^{2 + \left| q_i - q_j \right|}
        + X_{q_i} \delta_{ij} \\
        \lambda^{uR}_{ij} &\propto& \epsilon^{2 +
        \left| \bar{u}_i - \bar{u}_j \right|}
        + X_{u_i} \delta_{ij} \\
        \lambda^{dR}_{ij} &\propto& \epsilon^{2 +
        \left| \bar{d}_i - \bar{d}_j \right|}
        + X_{d_i} \delta_{ij}
\end{eqnarray}
The second term in each expression accounts for the $D$-term
contribution to masses, discussed above.

We have already argued that the up-type higgs $h_u$ must not carry
$X$-charge.  If the down-type higgs $h_d$ carries charge $X$,
a scalar $h_u h_d$ mass mixing or $\mu B$ term is induced by the operator
\be
  \frac{ \left( h_u h_d \right)_1 ( N^X )_1
  \left( P^* P \right)_{\theta \theta \bar\theta \bar\theta} }{ M^{2+X} } =
  \epsilon^{2+X} m^2 \phi_u \phi_d \ .
\ee
Furthermore, by eq.~(\ref{eq:sqmass}), the diagonal higgs' mass squared
$m^2_{h_u} \sim \epsilon^2 m^2$ and
$m^2_{h_d} \sim \left( X + \epsilon^2 \right) m^2$.
These results can be used in conjunction with the approximate relation
\be
  \label{tbeta}
  \tan\beta\sim {m^2_{h_u}+m^2_{h_d}\over \mu B}
\ee
which is valid for $\tan \beta \gtap 1$, to show that
\be
  \tan \beta \sim 1 \quad {\rm for} \, X = 0
  \qquad \qquad
  \tan \beta \sim \frac{1}{\epsilon^{2+X}} \quad {\rm for} \, X > 0 \ .
\ee
When neither higgs carries $X$-charge, $\tan \beta \sim 1$, and
the small values of $m_b / m_t$ and $m_{\tau} / m_t$ must be
explained by assigning nonzero $X$-charge to $\bar{d}_3$ and to
$\ell_3$ or $\bar{e}_3$.  Although the corresponding scalars will then
acquire masses of order $m$,
a heavy right-handed bottom squark and heavy third generation sleptons
do not upset the naturalness of the electroweak scale, because these
particles couple only weakly to higgses.
When $h_d$ carries $X$-charge, $\tan \beta$ is large\footnote{
The authors of \cite{NR} argued that $\tan \beta$ could never
be naturally large in a model with only two higgs doublets.
Our conclusions here do not contradict their result, because the
presence of multiple scales in the soft supersymmetry breaking
terms violates one of their assumptions.}, and the low-energy
effective theory below the scale $m$ contains only a single
higgs scalar.
Since the couplings of down-type quarks to $h_d$ are suppressed by
at least $X$ factors of $\epsilon$, the bottom to top mass ratio
at short distances is at most $\epsilon^{2+2X}$ in such scenarios.
Since $m_b / m_t \sim \epsilon^3$ is the
smallest ratio consistent with experiment, we are forced to reject
models with any $X$-charge in the higgs sector\footnote{
Supergravity gives a contribution to $\mu B \sim \epsilon^2 m^2$ when
$M$ is the Planck scale and $X=1$.  This implies $\tan \beta \sim 1 /
\epsilon^2$, allowing us to
obtain $m_b / m_t \sim m_{\tau} / m_t \sim \epsilon^3$,
provided no third generation superfield carries $X$-charge.
Although such models are interesting because of this constraint
and because the corresponding low-energy effective theories have only a
single higgs, we have been unable to find a set of $X$-charge assignments
of this type which suppresses FCNC sufficiently to be phenomenologically
viable.}.

Gaugino mass terms also arise as nonrenormalizable operators.  In particular,
terms of the form
\be
        \frac{ \left( W_{\mu} W^{\mu} \right)_1
        \left( N \right)_1 \left( P \right)_{\theta \theta} }{ M^2 } =
        \epsilon^2 m \Psi_{\lambda} \Psi_{\lambda}
\ee
induce gaugino masses of order $\epsilon^2 m \sim 200$~GeV.
A higgsino mass term of the same order is induced by the operator
\cite{GM}
\be
        \frac{ \left( h_u h_d \right)_{\theta \theta}
        \left( N^* \right)_1
        \left( P^* \right)_{\bar{\theta} \bar{\theta}} }{ M^2 } =
        \epsilon^2 m \Psi_{h_u} \Psi_{h_d} \ .
\ee
Gauginos and higgsinos are the lightest sparticles in these models.

We should note that there are numerous superpotential terms allowed
by all symmetry considerations which we nonetheless reject because
they give $M$-scale masses to MSSM fields.  Such ``dangerous'' terms
include, for example, $h_u h_d P N / M$, which gives
$\mu B \sim \epsilon^2 m M$.
The possibility of dangerous superpotential terms is a familiar
problem for supersymmetric models containing fields with vevs at high
scales.  Here we merely reiterate that, although we cannot forbid such terms
on symmetry grounds, the nonrenormalization theorem guarantees that
they will not be generated if absent initially.  Setting dangerous
terms equal to zero is thus at least technically natural.  There are no
dangerous K\"{a}hler potential terms, which is fortunate, because
the K\"{a}hler potential is not protected by any nonrenormalization
theorem.

A strong constraint on the values of $X$-charges
can be obtained by noting \cite{DG}
that integrating out matter above the scale $m$ to produce a low-energy
effective theory introduces an effective Fayet-Iliopoulos term for hypercharge
\be
 \frac{g_1^2 m^2}{( 4 \pi )^2} \left[
 {\rm tr} \left( X Y \right) \ln \left( \frac{M^2}{m^2} \right) -
 {\rm tr} \left( X Y \ln X \right) \right]
\ee
which can lead to disasterous color and electric charge breaking
minima of the scalar potential.  Because of the large log
in the first term\footnote{The calculation of the first term in this
expression
can and  should be improved by using the renormalization group, however its
magnitude and our conclusions are not affected.}  we are forced to require
\begin{equation} \label{eq:trace}{\rm tr}( X Y ) \sim 0 \ .\end{equation}
Since this trace requirement involves all matter
superfields in the theory, it prevents us from considering quarks and leptons
separately.  It can be accommodated nicely if we assign $X$-charges to $SU(5)$
multiplets, i.e. $\ell_i = \bar{d}_i$ and $q_i = \bar{u}_i = \bar{e}_i$.
Unfortunately, such an assignment predicts
\be
        \frac{ m_e }{ m_{\mu} } \sim \frac{ m_d }{ m_s } \qquad
        \frac{ m_{\mu} }{ m_{\tau} } \sim \frac{ m_s }{ m_b }
\ee
the first of which is off by an order of magnitude.  This problem,
which plagues all $SU(5)$-respecting  models,
has been addressed by Georgi and Jarlskog \cite{GJ}, who  showed how
$SU(5)$ group theory
factors can give the successful prediction
\be
\frac{ m_e }{ m_{\mu} } \sim  \frac{ m_d }{  9 m_s } \ .
\ee

If the  mixed standard model--$U(1)_X$ anomalies are cancelled entirely
by the Green--Schwarz mechanism, and if $\sin^2\theta_W=3/8$ at short
distances, then the $X$-charges must satisfy the
constraints (hereafter referred to as the GS constraints)
\be
  {\rm tr} \left( X T_a T_b \right) \propto
  {\rm tr} \left(  T_a T_b \right)
  \qquad
  {\rm tr} \left( X^2 Y \right) = 0
\ee
where $T_a$ are  generators of SM gauge group transformations.
These are   automatically satisfied if $X$-charge assignments respect
$SU(5)$ symmetry, and we do not give any examples which satisfy the GS
constraints which are not  consistent with $SU(5)$.
Since additional, superheavy matter fields (which might, for example,
get mass from the vev of $\phi_N$) could also contribute to the mixed
anomalies, we will not impose the GS constraints.

In table \ref{charges}, we  give  the best
examples of  $X$-charge assignments
which approximately reproduce all CKM matrix elements and known fermion mass
ratios,
and which satisfy the hypercharge trace constraint eq.~(\ref{eq:trace}). We
also give an example  which is
consistent with SU(5) symmetry, but in which several fermion mass ratios
are off by factors of up to ten.
For all models, we have assumed $\epsilon \sim 1/5$, $h_u = h_d = 0$, and
$\tan \beta \sim 1$.
We mark  the model consistent with $SU(5)$ symmetry with an asterix.

We will now proceed to derive the
implications of our charge assignments for FCNC and proton decay rates.

Given the quark and squark mass matrices derived above, we can
determine the intergenerational mixing matrix elements which appear
at quark-squark-gluino vertices; these are relevant to the computation
of the supersymmetric contributions to FCNC amplitudes.
If the similarity transformation $V_L^{\dag} \lambda^q V_R$ diagonalizes
a quark mass matrix, then the left- and right-handed squark mass matrices
in the basis of quark mass diagonalizing superfields are
\be
        \overline{\lambda}^{\tilde{q} L} =
        V_L^{\dag} \lambda^{\tilde{q} L} V_L  \qquad
        \overline{\lambda}^{\tilde{q} R} =
        V_R^{\dag} \lambda^{\tilde{q} R} V_R\ .
\ee
A short algebraic exercise will show that, provided the $i$th and $j$th
generation squarks are degenerate, i.e. carry the same $X$-charge,
the order-of-magnitude of the off-diagonal element
$\lambda^{\tilde{q}}_{ij}$ is unaffected by this transformation.
If, on the other hand, the squarks are nondegenerate, the the
order of magnitude of an off-diagonal element is
\be
        \overline{\lambda}^{\tilde{q}}_{ij} = \max
        \left( \lambda^{\tilde{q}} , V_{ij}   \right)\ .
\ee
The quark-squark-gluino mixing matrices are just the matrices $Z$
which diagonalize $Z^{\dag} \overline{\lambda}^{\tilde{q}} Z$.
The $\epsilon$-dependence of their entries can be determined by
perturbative diagonalization.

For example, in all our models, $\lambda^{\tilde{d} L}$ and $V^d_L$
have the form
\be
  \lambda^{\tilde{d}L} = \left(
  \begin{array}{ccc}
    3 & \, \epsilon^3 \, & \epsilon^5 \\
    \epsilon^3 & \, 2 \, & \epsilon^4 \\
    \epsilon^5 & \, \epsilon^4 \, & \epsilon^2
  \end{array} \right)
  \qquad
  V^d_L = \left(
  \begin{array}{ccc}
    1 & \, \epsilon \, & \epsilon^3 \\
    \epsilon & \, 1 \, & \epsilon^2 \\
    \epsilon^3 & \, \epsilon^2 \, & 1
  \end{array} \right) \ .
\ee
This implies
\be
  \bar{\lambda}^{\tilde{d}L} =
  \left( V^d_L \right)^{\dag} \lambda^{\tilde{d}L} \, V^d_L =
  \left( \begin{array}{ccc}
    3 & \, \epsilon \, & \epsilon^3 \\
    \epsilon & \, 2 \, & \epsilon^2 \\
    \epsilon^3 & \, \epsilon^2 \, & \epsilon^2
  \end{array} \right)
\ee
which is diagonalized by the left-handed down-type quark-squark-gluino
mixing matrix
\be
\label{ZdL}
  Z^{\tilde{d}L} = \left(
  \begin{array}{ccc}
    1 & \, \epsilon \, & \epsilon^3 \\
    \epsilon & \, 1 \, & \epsilon^2 \\
    \epsilon^3 & \, \epsilon^2 \, & 1
  \end{array} \right) \ .
\ee
Although equation (\ref{ZdL}) only indicates the leading order scaling
with $\epsilon$ of each matrix element, we can actually derive additional
constraints on the matrix.
For instance, since unitarity requires the first and second columns to
be orthogonal and $Z^*_{31} Z_{32}$ contributes to the inner product of
the columns only at ${\cal O}(\epsilon^5)$, we have
\be
  Z^{\tilde{d}L *}_{21} = - Z^{\tilde{d}L}_{12} + {\cal O}( \epsilon^3 )
\ee
Such constraints from unitarity will prove crucial in our calculations
of squark contributions to FCNC, to which we now proceed.

Several groups \cite{HKT,GGMS} have calculated the squark-gluino box
contributions to FCNC
for arbitrary squark masses and quark-squark-gluino mixing matrices.
In particular,
their contribution to the $K_L - K_S$ mass difference $\Delta m_K$,
whose measured value places stringent limits on any FCNC beyond the
SM, is
\begin{eqnarray}
\label{FCNC}
        \frac{ \Delta m_K }{ m_K } &=&
        \alpha_S^2 ( m ) \, f_K^2 \Bigg[
        \, \frac{11}{54} \,
        \frac{ Z^{dL*}_{1i} Z^{dL}_{2i} Z^{dL^*}_{1j} Z^{dL}_{2j} }{
        m_{Li}^2 - m_{Lj}^2 } \ln \left( \frac{ m_{Li}^2 }{ m_{Lj}^2 }
        \right) + \\ & &
        \, \frac{11}{54} \,
        \frac{ Z^{dR*}_{1i} Z^{dR}_{2i} Z^{dR^*}_{1j} Z^{dR}_{2j} }{
        m_{Ri}^2 - m_{Rj}^2 } \ln \left( \frac{ m_{Ri}^2 }{ m_{Rj}^2 }
        \right) + \nonumber \\ & &
        \, \left( \frac{1}{9} - \frac{2}{27} \, R \right) \,
        \frac{ Z^{dL*}_{1i} Z^{dL}_{2i} Z^{dR^*}_{1j} Z^{dR}_{2j} }{
        m_{Li}^2 - m_{Rj}^2 } \ln \left( \frac{ m_{Li}^2 }{ m_{Rj}^2 }
        \right) \, \Bigg] \nonumber
\end{eqnarray}
where
\be
\label{Rdef}
        R = \left( \frac{ m_K }{ m_s + m_d } \right)^2 \sim 10\ .
\ee
This expression relies on vacuum insertion and PCAC
to obtain matrix elements of quark
operators between $K^0$ and $\bar{K}^0$, and the approximation that
$m_{\tilde{g}} \ll m_{\tilde{q}}$.  Corrections to this expression
from nonzero gaugino masses are less than 10\%.  Although in principle
left-right squark mixing exists and contributes to FCNC, not only are
these mixing angles themselves smaller than those amongst left- and
right-handed squarks separately,
but the contributions of these mixings to $\Delta m_K$
are also down relative to those included above by
$(m_{\tilde{g}} / m_{\tilde{q}})^2$.
The full expression for $\Delta m_K$ with arbitrary
gluino mass and squark mixing is relegated to an appendix.
For each set of charge assignments, we have computed the minimum
value of $m$ which will just suppress $\Delta m_K$ to its
observed size, and listed the result in table \ref{charges}.

We illustrate the calculation of $m_{\rm min}$ by computing the contribution
of the first and second generation left-handed down squarks to
expression (\ref{FCNC}).  Using the unitarity constraint relating
$Z^{\tilde{d}L}_{12}$ to $Z^{\tilde{d}L}_{21}$ derived earlier,
we obtain a result to leading order in $\epsilon$ proportional to
\be
  \left( Z^{\tilde{d}L}_{21} \right)^2
  \left[ \frac{1}{X_1} + \frac{1}{X_2} -
  \frac{2}{X_1 - X_2} \ln \left( \frac{X_1}{X_2} \right) \right]
\ee
where $X_1$ and $X_2$ are the $X$-charges of the first and second
generation left-handed down-squarks.  For $X_1 = X_2$, the factor in
brackets vanishes, illustrating the squark degeneracy mechanism
of FCNC suppression.  For our nondegenerate $X$-charge assignments,
the expression reduces to
\be
  \epsilon^2 \left[ \frac{1}{3} + \frac{1}{2} -
  2 \ln \left( \frac{3}{2} \right) \right] \sim 0.022 \epsilon^2 \ .
\ee
This is much smaller than the $\sim \epsilon^2$ value one might
have expected for ``generically'' nondegenerate squark masses.

Similar arguments, combining relations among matrix elements imposed
by unitarity with the explicit formula (\ref{FCNC}), can be applied
to the remaining terms.
We find that the $X$-charge assignments chosen to reproduce
the observed quark and lepton mass ratios provide significantly
more suppression of FCNC than naive estimates for generically
nondegenerate squarks would suggest.
For charge sets A and C, the dominant supersymmetric contribution
to FCNC, and thus the strongest constraint on $m_{\rm min}$,
comes from box graphs containing a left-handed first or second
generation squark and a right-handed third generation squark.
For sets B and D, graphs with both left and right-handed first and second
generation squarks dominate.

In all cases $m_{\rm min}$ is low enough that the predicted masses of
sparticles coupled strongly to the higgs satisfy the 't Hooft
bound ($\epsilon m_{\rm min} \ltap 1$ TeV) arising from the assumption of
1-loop naturalness of the electroweak scale.

CP violation could still produce a complex phase for $\Delta m_K$ which
would give unacceptably large $\epsilon_K$.
For generic phases of ${\cal O}(\pi/2)$, suppressing the contribution to
$\epsilon_K$ to the observed level would require $m_{\rm min}$ values
roughly an order of magnitude larger.
Such a large values would give third generation sparticle masses which
imply a fine-tuning of the electroweak scale to better than one part in
$10^3$, in gross violation of the 't Hooft bound.
Instead, we will simply assume imaginary contributions to
$\Delta m_K$ to be small.
Although we have no explanation for the near-reality of the
supersymmetry breaking terms within
the context of our model, a more detailed  model incorporating
spontaneous CP violation could surely be found which would suppress
$\epsilon_K$ (see, for example, ref.~\cite{NirRat}).
Imposition of  CP symmetry at short distances is theoretically
attractive, since CP is a gauge symmetry in certain theories with
extra dimensions at the Planck scale, such as string theory~\cite{CPgauge}.

Another potential problem arises when we consider the requirement
of electroweak scale naturalness at two loops, where there
is a contribution to the higgs mass squared proportional to $m^2$
which is enhanced by a large logarithmic factor, $\ln(M^2/m^2)$.
Dimopoulos and Giudice \cite{DG} computed the effects of heavy
squark and slepton masses on the stability of the electroweak scale
using the 2-loop renormalization group equations, and
concluded that requiring that contributions to the
higgs mass squared should not have to cancel to better than 10\% implies
that all squarks and sleptons are lighter than $\sim$~2--5~TeV (DG
bound).
All of our models, with $m$ set to $m_{\rm min}$,
are in mild violation of this requirement.
Note, however, that, were we to allow the same 10\% fine tuning amongst the
squark contributions to $\Delta m_K$ that we have already allowed
in the higgs sector, each $m_{\rm min}$ would be about a factor of three
lower and all the models would satisfy the 2-loop constraint.
Alternatively, the DG bound could be relaxed by lowering the scale
$M$ of $U(1)_X$ breaking, which would reduce, in turn, the size of
the large logarithms which enhance loop corrections.

The possible appearance in the effective superpotential of dimension-5
operators of the form $qqq\ell$ and $\bar u\bar u \bar d \bar e$,
which generically induce proton decay at a
rate far above the experimental limits even when suppressed by the Planck
scale, has long posed a problem for
supersymmetric extensions of the standard model \cite{SY}.
The existence of $X$-charge suppresses such operators by allowing only
higher-dimension operators of the form
\be
        \frac{ N^n (qqq\ell,\bar u\bar d\bar d\bar e) }{ M^{n+1} }, \sim
        \frac{ \epsilon^n }{ M } (qqq\ell,\bar u\bar u\bar d\bar e)
\ee
where $n$ is chosen to form a $U(1)_X$ gauge-invariant operator. The most
severe constraint on our models comes from the operator
$q_1 q_1 q_2 \ell_3$ which causes a proton to decay into a kaon and a tau
neutrino. A simple
one-loop calculation for $m_p \ll m_{\tilde{g}} \ll m_{\tilde{q}}$
and naive dimensional analysis show that the rate is
\be
        \Gamma_p
        \sim
        \frac{ \alpha^2_S (m_{\tilde{q}}) }{ 4\left( 4 \pi \right)^5 }
        \frac{ m_{\tilde{g}}^2 m_p^5 \epsilon^{2n} }{ m_{\tilde{q}}^4 M^2}
        \sim\alpha^2_S (m)
        \frac{ \epsilon^{4+2n} }{4096  \pi^5}
        \frac{ m_p^5 }{ M^2 m^2 } \ .
\ee
Note that this expression for $\Gamma_p$ is depressed from the result for
the MSSM with a single SUSY scale not only by several powers of $\epsilon$,
but also by the factor $( m_{\tilde{g}} / m_{\tilde{q}} )^2$.
This additional suppression arises from a gluino mass insertion in the
relevant diagram.
Experimental limits on the proton lifetime require $n > 5$ for $m \sim 5$~TeV
and $M$ set equal to the Planck mass.
All $X$-charge assignments under consideration give a proton
lifetime of greater than $10^{40}$~years with $m = m_{\rm min}$ and a
Planck-scale $M$,
and are consistent with proton lifetime limits for any
$M \gtap 10^{15}$~GeV.
Dimension-6
proton-decay inducing operators will be similarly suppressed, but these
are not a phenomenological problem for $M$ set to the Planck scale.
Finally, we should note that the $U(1)_X$ model does not alleviate the
problem of $B$ and $L$ violating dimension-4 operators, which must still
be forbidden by imposing a symmetry such as $R$-parity.  Our
philosophy has been to avoid the imposition of global symmetries;
$R$-parity, however, could easily arise automatically as a consequence of
a spontaneously broken $B-L$ gauge symmetry~\cite{smartin}.

The horizontal, anomalous $U(1)$ gauge group models presented here
have many unusual and attractive features.  The charge assignments
considered reproduce the observed fermion mass hierarchy and
CKM mixing matrix elements.
These same charge assignments predict a most
unusual pattern of superpartner masses.  Squarks and sleptons are highly
nondegenerate, with a mass hierarchy that is the mirror-image of the
fermion mass hierarchy: generically light third generation and progressively
heavier second and first generation sparticles.
In particular, top and left-handed bottom squarks are predicted to
weigh in between 500 GeV -- 1 TeV.  For some charge assignments, some
other third generation sparticles are also found at this scale.
In other cases, they are found at the higher mass scale (2--10 TeV) of
the the first and second generation superpartners.
Gauginos and higgsinos are predicted to be the lightest (100--200 GeV)
superpartners.
We have shown that, despite their nondegeneracy, the sparticles make
acceptably small contributions to FCNC.
Also, dimension-5 proton decay amplitudes are suppressed sufficiently
to satisfy experimental bounds.
The low masses of third generation sparticles which couple strongly to
the higgs allow the the electroweak symmetry-breaking scale to be
reproduced with only mild fine-tuning.
On the downside, our models do not explain the smallness of observed CP
violation or the absence of certain dangerous terms generically allowed
in the superpotential.

Aside from these few but noteworthy unresolved mysteries, we have
successfully used a
horizontal, anomalous, broken $U(1)$ gauge symmetry to construct
several phenomenologically acceptable ``more'' minimally supersymmetric
extensions of the standard model.
Such models offer the tantalizing prospect of allowing us to one day
determine Froggatt-Nielsen charges experimentally by simply measuring
squark and slepton mass ratios. From a theoretical perspective,
perhaps the most attractive feature
of these models is their unification of the supersymmetry-breaking
and flavor physics sectors into a single sector with an uncomplicated
gauge group and small number of additional matter superfields.

\bigskip\noindent{\bf Acknowledgements}
After this work was completed, a complementary study, which consided the
case of a $U(1) \times U(1)$ horizontal gauge group \cite{zhang}, appeared.
One of the authors (D.W.) would like to thank Francois Lepeintre for
helpful discussions.
This work was supported by the U.S. Department of Energy,
grant DE-FG03-96ER40956.

\section{Appendix}

Previous studies have given formulae for SUSY contributions to FCNC
amplitudes only in the limit of
squark near-degeneracy \cite{GGMS} or without explicitly evaluating
various integrals which appear in the general case \cite{HKT}.  We
therefore present, for future reference, a formula\footnote{
The first three terms of this this formula agree with the analogous
terms of formula (II.1) of ref.~\cite{HKT}, provided the coefficient
$1/36$ in their formula is changed to $11/36$, a replacement also
necessary for consistency with formulas appearing later in their article.
The remaining three terms are in substansive disagreement with
ref.~\cite{HKT}.   All terms agree, in the appropriate limit, with the
calculation of ref.~\cite{GGMS}, and are supported by an independent
calculation \cite{FL}.}
for the squark-gluino box contribution to the $K^0 - \bar{K}^0$ mixing
amplitude valid for arbitrary masses and mixing angles.
\begin{eqnarray}
        \lefteqn{ \langle K^0 \left| H \right| \bar{K}^0 \rangle =
        \alpha_S^2 \, m_K \, f_K^2 \times}
        \\ & &
        \Bigg\{ \, M_1 \,
        \left( \frac{11}{36} A_{i,j} + \frac{1}{9} B_{i,j} \right) \,
        Z^{*}_{1L,i} Z^{}_{2L,i} Z^{*}_{1L,j} Z^{}_{2L,j}
        + \nonumber
        \\ & &
        \tilde{M_1} \,
        \left( \frac{11}{36} A_{i,j} + \frac{1}{9} B_{i,j} \right) \,
        Z^{*}_{1R,i} Z^{}_{2R,i} Z^{*}_{1R,j} Z^{}_{2R,j}
        + \nonumber \\ & &
        \left[ \left( \frac{5}{9} M_5 - \frac{1}{3} M_4 \right) A_{i,j}
        + \left( \frac{7}{3} M_4 + \frac{1}{9} M_5 \right) B_{i,j} \right] \,
        Z^{*}_{1L,i} Z^{}_{2L,i} Z^{*}_{1R,j} Z^{}_{2R,j}
        + \nonumber \\ & &
        \left( \frac{17}{18} M_2 - \frac{1}{6} M_3 \right) \,
        B_{i,j} \,
        Z^{*}_{1L,i} Z^{}_{2R,i} Z^{*}_{1L,j} Z^{}_{2R,j}
        + \nonumber \\ & &
        \left( \frac{17}{18} \tilde{M_2} - \frac{1}{6} \tilde{M_3} \right) \,
        B_{i,j} \,
        Z^{*}_{1R,i} Z^{}_{2L,i} Z^{*}_{1R,j} Z^{}_{2L,j}
        - \nonumber \\ & &
        \left( \frac{11}{18} M_4 + \frac{5}{6} M_5 \right) \,
        A_{i,j} \,
        Z^{*}_{1L,i} Z^{}_{2R,i} Z^{*}_{1R,j} Z^{}_{2L,j}
        \Bigg\}
\end{eqnarray}
Here $Z$ is the down-type quark-squark-gluino mixing matrix;
the first and second indices run over left- and right-handed down-type
quarks and squarks, respectively.
The $M$'s are matrix elements of four-quark operators between $K^0$ and
$\bar{K}^0$ states; their values from PCAC and vaccuum insertion are
listed for reference.
\begin{eqnarray}
  M_1 &=& \langle K^0 | \bar{d}^{\alpha}_L \gamma_{\mu} s^{\alpha}_L
  \bar{d}^{\beta}_L \gamma_{\mu} s^{\beta}_L | \bar{K}^0 \rangle =
  \frac{2}{3} \\
  M_2 &=& \langle K^0 |
  \bar{d}^{\alpha}_R s^{\alpha}_L \bar{d}^{\beta}_R s^{\beta}_L
  | \bar{K}^0 \rangle = - \frac{5}{24} R \\
  M_3 &=& \langle K^0 |
  \bar{d}^{\alpha}_R s^{\beta}_L \bar{d}^{\alpha}_R s^{\beta}_L
  | \bar{K}^0 \rangle = \frac{1}{12} R \\
  M_4 &=& \langle K^0 |
  \bar{d}^{\alpha}_R s^{\alpha}_L \bar{d}^{\beta}_L s^{\beta}_R
  | \bar{K}^0 \rangle = \frac{1}{12} + \frac{1}{2} R \\
  M_5 &=& \langle K^0 |
  \bar{d}^{\alpha}_R s^{\beta}_L \bar{d}^{\beta}_L s^{\alpha}_R
  | \bar{K}^0 \rangle = \frac{1}{4} + \frac{1}{6} R
\end{eqnarray}
Expressions for $\tilde{M}$'s are obtained from those for the corresponding
$M$'s by the exchange $L \leftrightarrow R$.  The ratio $R$ is defined by
eq.~\ref{Rdef}.
The functions $A_{i,j}$ and $B_{i,j}$ depend on the masses of the
$i$th and $j$th down-type squark and the gluino mass and have the
explicit form
\begin{eqnarray}
  A_{i,j} &=&
  \frac{ m^2_{\tilde{g}} }{ \left( m_i^2 - m^2_{\tilde{g}}
  \right) \left( m_j^2 - m^2_{\tilde{g}} \right) } +
  \frac{ m_i^4 }{ \left( m_i^2 - m_j^2 \right) \left( m_i^2 - m^2_{\tilde{g}}
  \right)^2 } \ln \left( \frac{ m_i^2 }{ m^2_{\tilde{g}} } \right) +
  \nonumber
\\ & &
  \frac{ m_j^4 }{ \left( m_j^2 - m_i^2 \right) \left( m_j^2 - m^2_{\tilde{g}}
  \right)^2 } \ln \left( \frac{ m_j^2 }{ m^2_{\tilde{g}} } \right)
  \\
  B_{i,j}
&=&
  \frac{ m^2_{\tilde{g}} }{ \left( m_i^2 - m^2_{\tilde{g}}
  \right) \left( m_j^2 - m^2_{\tilde{g}} \right) } +
  \frac{ m_i^2 m_{\tilde{g}}^2 }{ \left( m_i^2 - m_j^2 \right)
  \left( m_i^2 - m^2_{\tilde{g}}
  \right)^2 } \ln \left( \frac{ m_i^2 }{ m^2_{\tilde{g}} } \right) +
  \nonumber
\\ & &
  \frac{ m_j^2 m_{\tilde{g}}^2 }{ \left( m_j^2 - m_i^2 \right)
  \left( m_j^2 - m^2_{\tilde{g}}
  \right)^2 } \ln \left( \frac{ m_j^2 }{ m^2_{\tilde{g}} } \right)
\end{eqnarray}
Equation \ref{FCNC} is obtained by ignoring mixing between left- and
right-handed squarks, assuming PCAC and vaccuum insertion values for
the relevant matric elements, and evaluating $A$ and $B$
in the $m_{\tilde{g}} \rightarrow 0$ limit.  Formula (4) of
ref.~\cite{GGMS} for SUSY contributions to FCNC in the limit of
nearly degenerate squarks may be obtained by setting
$m_i^2 = m_j^2 + \delta m_{ij}^2$ and expanding $A$ and $B$ up
to terms quadratic in $\delta m_{ij}^2$.




\begin{table}
\label{charges}
\vspace{1cm}
\begin{tabular}{|l|c|c|c|c|c|c|c|}
 set & $q_1,q_2,q_3$ & $\bar u_1,\bar u_2,\bar u_3$ &
 $\bar d_1,\bar d_2,\bar d_3$ & $\ell_1,\ell_2,\ell_3$ &
 $\bar e_1,\bar e_2,\bar e_3$ & $h_u,h_d$ &
 $m_{\rm min} {\rm (TeV)}$  \\ \hline
 A     & 3,2,0 & 4,1,0 & 3,3,2 & 3,3,2 & 4,1,0 & 0,0 & 4.2 \\ \hline
 B     & 3,2,0 & 4,1,0 & 4,3,2 & 4,3,2 & 4,1,0 & 0,0 & 3.5 \\ \hline
 C     & 3,2,0 & 4,1,0 & 3,3,2 & 4,3,1 & 3,1,1 & 0,0 & 4.2 \\ \hline
 D$^*$ & 3,2,0 & 3,2,0 & 4,3,2 & 4,3,2 & 3,2,0 & 0,0 & 3.5
\end{tabular}
\vspace{1cm}
\caption{X-charge assignments for left chiral superfields. The column
$m_{\rm min}$ gives the minimum value for $m$ consistent with FCNC.
All models are also consistent with limits on proton decay.
Set D is consistent with SU(5) symmetry and with the GS constraints. }
\end{table}


\end{document}